\documentclass{iau}
\usepackage{graphicx}

\newcommand{\Mpc}{\,$h^{-1}$\,Mpc}
\newcommand{\zeff}{z_{\rm eff}}

\title[IAUS289.~~H$_{\it 0}$ from the 6dF Galaxy Survey] %% short title %%
{Measuring H$_{\bf 0}$ from the 6dF Galaxy Survey and future low-redshift surveys}

\author[M. Colless, F. Beutler, \& C. Blake]
{Matthew Colless,$^1$ Florian Beutler,$^{2,3}$ \and Chris Blake$^4$}
\affiliation{$^1$Australian Astronomical Observatory,
  P. O. Box 915, North Ryde, NSW 1670, Australia \\
  email: {\tt colless@aao.gov.au} \\[\affilskip]
  $^2$International Centre for Radio Astronomy Research, University of
  Western Australia,\\ 35 Stirling Highway, Perth, WA 6009, Australia \\[\affilskip]
  $^3$Lawrence Berkeley National Laboratory, 1 Cyclotron Road, Berkeley,
  CA 94720, USA \\
  email: {\tt fbeutler@lbl.gov} \\[\affilskip]
  $^4$Centre for Astrophysics \& Supercomputing, Swinburne University of
  Technology,\\ P. O. Box 218, Hawthorn, VIC 3122, Australia \\ 
  email: {\tt cblake@astro.swin.edu.au}}

\pubyear{2012}
\volume{289}  %% IAU Symposium No.
\jname{Advancing the Physics of Cosmic Distances}
\editors{Richard de Grijs \& Giuseppe Bono, eds.}

\begin{document}

\maketitle

\begin{abstract}
  Baryon acoustic oscillations (BAO) at low redshift provide a precise
  and largely model-independent way to measure the Hubble constant,
  H$_0$. The 6dF Galaxy Survey measurement of the BAO scale gives a
  value of H$_0 = 67 \pm 3.2$ km s$^{-1}$ Mpc$^{-1}$, achieving a
  $1\sigma$ precision of 5\%. With improved analysis techniques, the
  planned {\sc wallaby} (H{\sc i}) and {\sc taipan} (optical) redshift
  surveys are predicted to measure H$_0$ to 1--3\% precision.
\end{abstract}

%\index[author]{Colless, M.|textbf}
%\index[author]{Beutler, F.}
%\index[author]{Blake, C.}

%\index[subject]{cosmology!observations}
%\index[subject]{surveys}
%\index[subject]{distance scale}
%\index[subject]{large-scale structure of universe}

\firstsection 
\section{Introduction}

Baryon acoustic oscillations (BAO) produced by the interaction of
photons and baryons in the early universe provide an absolute standard
rod that is calibrated by observations of the cosmic microwave
background (CMB). The BAO scale is determined by well-understood linear
physics and depends only on the physical densities of dark matter and
baryons. In principle, the BAO scale can be measured to about 1\%
precision from tracers of the large-scale structure of the universe over
a wide range of redshifts. It is therefore a powerful probe of cosmic
geometry (\cite{Seo2003}; \cite{Blake2003}), particularly as it can be
used to measure the evolution of both the Hubble parameter $H(z)$
radially along the line of sight and the angular diameter distance
$D_A(z)$ tangentially across the line of sight. However to achieve the
full precision possible from the BAO scale requires large samples of
tracers ($\sim$10$^6$ objects) over large volumes ($\sim$1\,Gpc$^3$).

BAO are complementary to other probes of the universe's geometry, such
as supernova measurements of luminosity distance $D_L(z)$, in that they
measure different cosmological properties and have a different physical
basis (and therefore have different sources of systematic errors). The
main potential sources of systematic errors for BAO measurements are
non-linear clustering, redshift-space distortions and possible
scale-dependent bias.

At low redshift, BAO yield a measurement of the distance scale that
requires only the CMB calibration of the sound horizon scale and is
largely independent of the details of the cosmological model
(\cite{Beutler2011}). Thus a measurement of the BAO scale in a
low-redshift ($\bar{z} \approx 0.05$) galaxy survey like the
6dF Galaxy Survey (6dFGS) yields a direct and nearly model-independent
measurement of the Hubble constant $H_0$.

\section{The 6dF Galaxy Survey}

The 6dF Galaxy Survey (6dFGS) is a redshift and peculiar velocity survey
of the southern sky (Jones \etal\ 2004, 2006, 2009). It used the 6dF
multi-fibre spectrograph on the UK Schmidt Telescope (UKST) operated by
the Australian Astronomical Observatory (AAO) to spectroscopically
survey a sample of near-infrared galaxies selected from the 2MASS
Extragalactic Source Catalog (XSC; \cite{Jarrett2000}) covering the
whole southern sky outside of 10$^\circ$ from the Galactic plane. The
6dFGS measured redshifts for more than 125000 galaxies with $K \leq
12.65$ and Fundamental Plane peculiar velocities for about 9000
early-type galaxies. The results discussed here are based on the 6dFGS
redshift survey, for which the median redshift is $z=0.052$.

For the purpose of measuring the correlation function, galaxies were
excluded from the sample if they lay in sky regions with completeness
$<$60\%. This reduced the sample to 75117 galaxies. The selection
function was derived by scaling the survey completeness as a function of
magnitude to match the integrated on-sky completeness using mean galaxy
counts. The effective weighted volume of the sample is
0.08\,$h^{-3}$\,Gpc$^3$, and the effective redshift at which the BAO
scale is measured is $\zeff=0.106$.

\firstsection 
\section{The galaxy correlation function and $H_0$}

\begin{figure}[t]
\begin{center}
 \includegraphics[width=10cm]{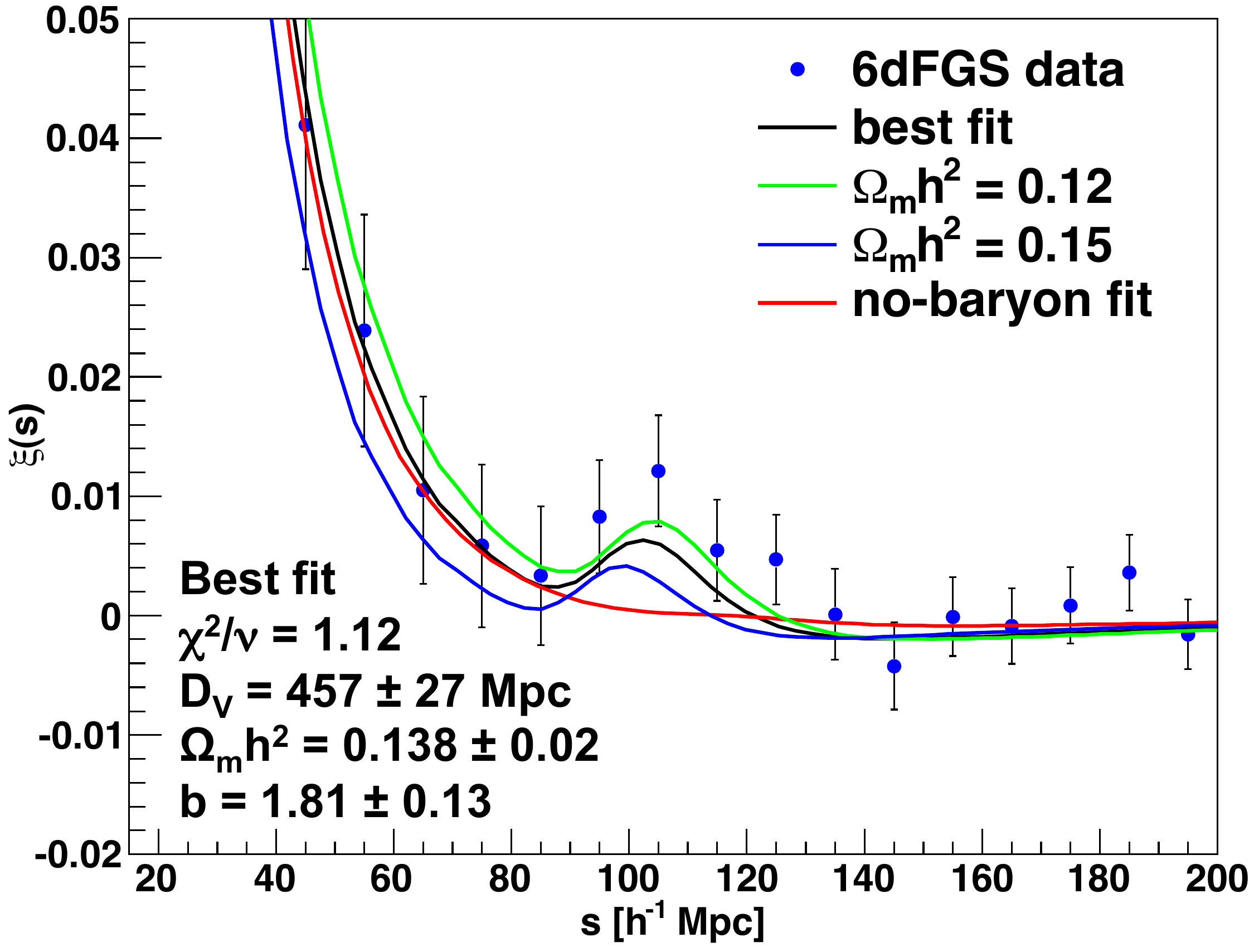} 
 \caption{The BAO signal in the 6dFGS correlation function. The measured
   correlation function and errors are shown as blue dots with
   errorbars. The best-fit model (black curve) is also shown, along with
   two flanking models (blue and green curves) and the best-fit
   no-baryon model (red curve). The parameters of the best-fit model are
   given in the legend at lower left.}
   \label{fig1}
\end{center}
\end{figure}

Figure~\ref{fig1} shows the correlation function for this sample of
galaxies computed using the \cite{Landy1993} method with inverse density
weighting following \cite{Feldman1994} and an integral constraint
correction. The errorbars on the correlation function points are based
on log-normal realisations. Full details of the methodology are given in
\cite{Beutler2011}. The BAO peak in the correlation function at a scale
of 105\Mpc\ is clearly visible.

The correlation function is modelled accounting for the wide-angle
effects in this large-area survey, the effects of non-linear evolution
in the galaxy clustering, and the scale-dependence of the bias; again,
details are given in \cite{Beutler2011}. The 6dFGS sample is not large
enough to constrain $H(\zeff)$ and $D_A(\zeff)$ separately using the 2D
correlation function; instead we constrain the combined quantity
$D_V(\zeff)$ (\cite{Eisenstein2005}) using the 1D correlation function.

The model for the correlation function uses parameter values from WMAP7
(\cite{Komatsu2010}) to define the power spectrum and the BAO scale. We
fit the correlation function over the range 10\Mpc\ to 190\Mpc. The free
parameters in our model are the physical matter density $\Omega_{m}h^2$,
the bias $b$, the non-linear damping scale $k_*$, and the scale
distortion parameter $\alpha = D_v(\zeff)/D_v^{\rm fid}(\zeff)$ that
measures the deviation of the BAO scale from the fiducial cosmological
model.

\begin{figure}[t]
\begin{center}
 \includegraphics[width=10cm]{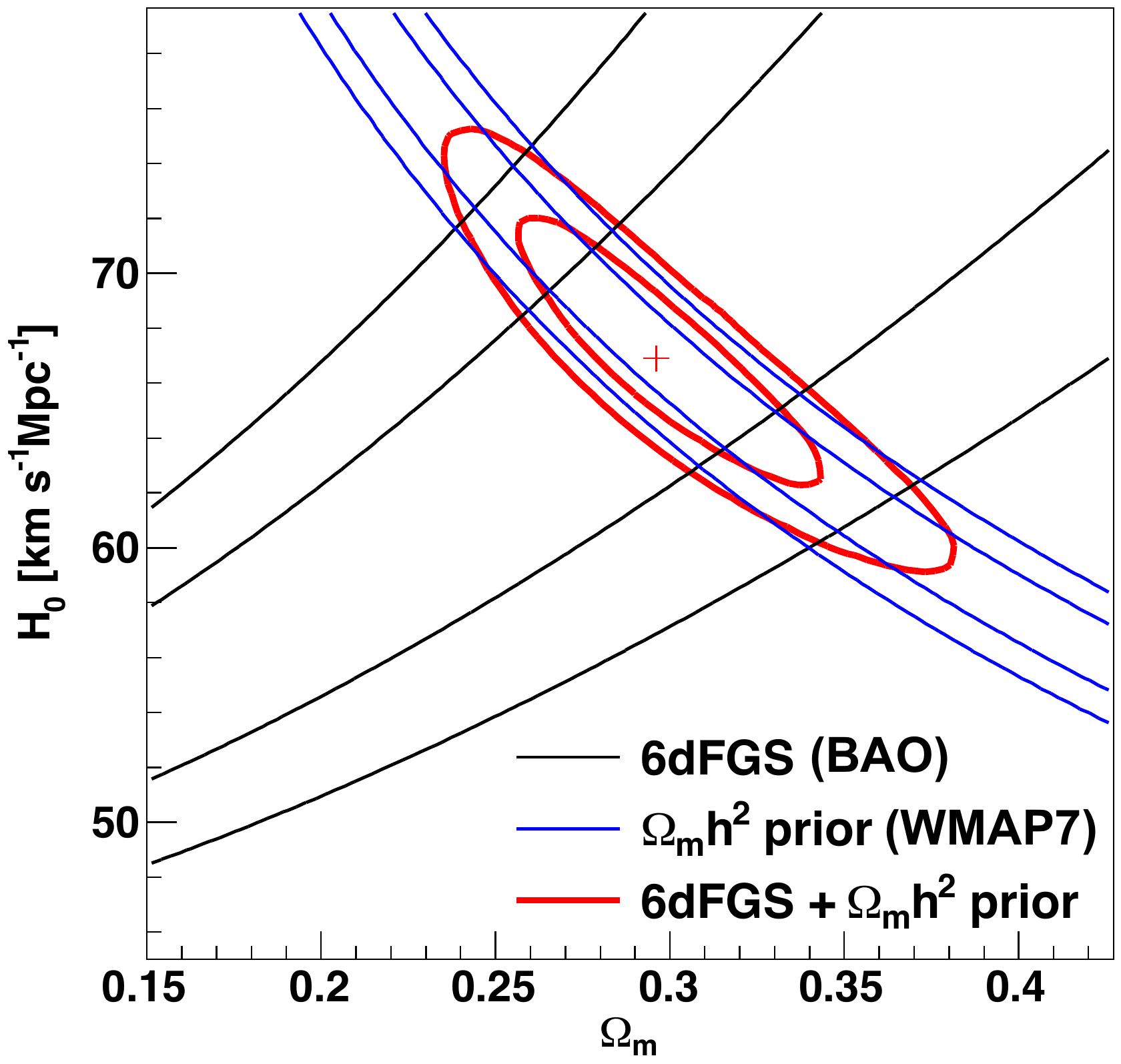} 
 \caption{The constraints on $H_0$ and $\Omega_m$ from the 6dFGS BAO
   measurement (black contours), the WMAP7 CMB observations (blue) and
   the BAO and CMB measurements combined (red).}
   \label{fig2}
\end{center}
\end{figure}

Our best-fit model for the correlation function is shown in
Figure~\ref{fig1} and has parameters $\Omega_{m}h^2 = 0.135 \pm 0.020$,
$b = 1.65 \pm 0.10$, $k_* > 0.19\,h\,{\rm Mpc}^{-1}$ (95\% confidence
lower limit), and $\alpha = 1.039 \pm 0.062$. This corresponds to
$D_V(\zeff) = 457 \pm 27$~Mpc, with a precision of 5.9\%. Marginalising
over $k_*$ and using a prior on $\Omega_{m}h^2$ from WMAP7, we obtain a
measurement for the Hubble constant of $H_0 = 67 \pm
3.2$\,km\,s$^{-1}$\,Mpc$^{-1}$, which has an uncertainty of 4.8\%. The
corresponding estimate for $\Omega_m$ is $0.296 \pm 0.028$. The joint
constraints on $H_0$ and $\Omega_m$ are shown in Figure~\ref{fig2}.

The 6dFGS result for $H_0$ (\cite{Beutler2011}) has comparable precision
to that published recently from the SH0ES distance-ladder program
(\cite{Riess2011}), $H_0 = 73.8 \pm 2.4$\,km\,s$^{-1}$\,Mpc$^{-1}$,
although our result is 1.7$\sigma$ lower. It is similarly comparable to,
and in better agreement with, the model-dependent WMAP7 estimate
(\cite{Komatsu2010}), $H_0 = 70.3 \pm 2.5$\,km\,s$^{-1}$\,Mpc$^{-1}$.
The advantages of the 6dFGS result are that it is not reliant on a
series of distance ladder steps (unlike the SH0ES result) and that it is
largely independent of the cosmological model (unlike the WMAP7 result).

\section{Constraints on $H_0$ from future surveys}

How precise could future $H_0$ measurements from low-redshift galaxy
surveys be? Two large low-$z$ galaxy surveys are in prospect: the
WALLABY HI survey (\cite{Duffy2012}) planned for the Australian SKA
Pathfinder and the TAIPAN optical survey planned for the UKST
(\cite{Beutler2011}). WALLABY is expected to start in 2014--15. It will
cover the entire sky (with a matching Westerbork survey) and give
redshifts for $\sim$6$\times$10$^5$ galaxies with $b \approx 0.7$ and
$\bar{z} \approx 0.04$ over a volume $V_{\rm eff} \approx
0.12\,h^{-3}$\,Gpc$^3$. TAIPAN is a southern-sky survey expected to
start in 2015. It will (at a limit of $r< 17$) give redshifts for
$\sim$4$\times$10$^5$ galaxies with $b \approx 1.6$ and $\bar{z} \approx
0.07$ over a volume $V_{\rm eff} \approx 0.23\,h^{-3}$\,Gpc$^3$.

\begin{figure}[t]
\begin{center}
 \includegraphics[width=32pc]{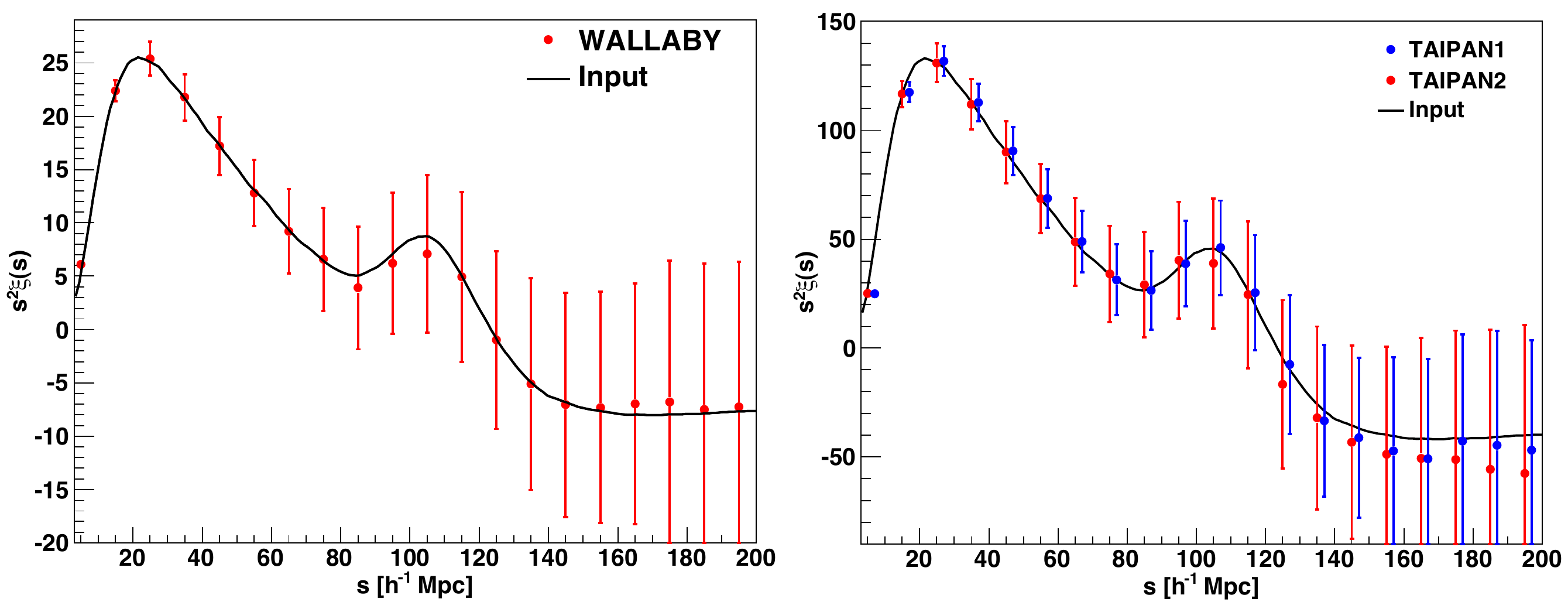} 
 \caption{Predicted BAO signals and uncertainties in the galaxy
   correlation function from the WALLABY HI survey with ASKAP (left) and
   the TAIPAN survey with UKST (right).}
   \label{fig3}
\end{center}
\end{figure}

Figure~\ref{fig3} shows the predicted BAO signal in the correlation
functions for both these surveys, based on 100 log-normal realisations.
We find that WALLABY obtains essentially the same precision in measuring
$H_0$ as 6dFGS. However the deeper TAIPAN survey, with its larger
effective volume and higher bias, can measure $H_0$ with 3\% precision.
In addition, density field reconstruction has shown significant
improvement in the cosmological parameter constraints by using extra
information from the density field (\cite{Padmanabhan2012}). At low
redshift this gives an improvement of about a factor of two, so could
improve the precision of the $H_0$ measurement for 6dFGS and WALLABY to
$\sim$2.5\% and for TAIPAN to $\sim$1.5\%. A combined analysis using
both the low-bias WALLABY galaxies and the high-bias TAIPAN galaxies
could do even better.

\firstsection

%\begin{discussion}
%\end{discussion}

\end{document}